\documentclass[a4paper,12pt,oneside,final,openbib]{article}
\pdfoutput=1
\usepackage[utf8]{inputenc}
\usepackage{url}
\usepackage{cite}
\usepackage{hyperref}
\usepackage{breakurl}
\usepackage{graphicx}
\usepackage{amsmath,amssymb,epsfig}
\usepackage{color}
\usepackage{mathrsfs}
\usepackage{subcaption}
\usepackage{mathtools}
\usepackage{slashed}

\newcommand{\MSB}{\overline{\textrm{MS}}}
\newcommand{\corresponds}{\mathrel{\widehat{=}}}
\newcommand{\dLSZ}{\delta\hspace{-0.6mm}\mathscr{Z}}

\begin{document}

\title{
{\hfill \normalsize MITP/23-058} 
\\ 
\vspace*{48pt}
\Large\bf 
Matching the Weak Mixing Angle 
at Low Energies 
\\[3ex] 
}
\author{\large  
Hubert Spiesberger, Stephan Wezorke 
\\ 
{\normalsize \em 
PRISMA$^+$ Cluster of Excellence,} 
\\
{\normalsize \em 
Institute for Nuclear Physics and Institute of Physics,} 
\\
{\normalsize \em 
Johannes Gutenberg-University, 55099 Mainz, Germany} 
} 

\maketitle

\begin{abstract}
The running weak mixing angle is used as a convenient tool 
to keep control of dominating logarithmic corrections in 
the Standard Model of the electroweak interactions connecting 
measurements at largely differing energy scales. To relate the 
solutions of the renormalization group equation for the weak 
mixing angle above and below particle thresholds one needs 
matching conditions. We suggest to define the 
matching at the $W$-boson threshold by comparing the full 
Standard Model with a low-energy effective theory where 
the weak neutral current, in particular for parity-violating 
fermion scattering, is described by 4-fermion operators 
and calculate the corresponding matching relation. 
\end{abstract}

\newpage

\section{Motivation and Introduction} 

A central parameter of the Standard Model (SM) of the 
electroweak interactions is the weak mixing angle, 
$\sin^2\theta_W$. Presently, its most precise determinations 
come from experiments at high energies, i.e.\ from $e^+ e^-$ 
annihilation at LEP on the $Z$-resonance and 
from Drell-Yan production in $\bar{p}p$ or $pp$ scattering 
at the Tevatron or the LHC, resp., also at the $Z$-pole 
\cite{Workman:2022ynf}. Less precise measurements at low 
energies are available from parity-violating atomic 
transitions or from elastic scattering of polarized 
electrons off protons at the QWeak experiment 
\cite{Qweak:2018tjf}. Upcoming experiments like 
P2 \cite{Becker:2018ggl} and MOLLER \cite{MOLLER:2014iki} with 
scattering of polarized electrons off protons or electrons 
are expected to provide measurements of $\sin^2\theta_W$ 
with a precision which is as precise as the high-energy 
measurements. 

A convenient way to compare low- and high-energy measurements 
is to use the running weak mixing angle. This can be 
defined either as an effective parameter which absorbs 
certain momentum-dependent higher-order corrections 
\cite{Czarnecki:1998xc,Ferroglia:2003wa} or in a framework 
with $\MSB$ renormalization as the solution of a renormalization 
group equation (RGE), often denoted $\sin^2\hat{\theta}_W(\mu)$ 
\cite{Erler:2004in,Erler:2017knj}. In both cases, the 
running weak mixing angle is a book-keeping quantity 
which allows one to describe the dominant logarithmic 
dependence on the typical energy of the process.  

At high energy scales, the RGE is governed by the complete 
particle content of the SM, including the $W^\pm$- and 
Higgs bosons and the heavy quarks. At low energy scales, 
the running of the weak mixing angle should, however, be 
determined by those particles whose masses are in the 
considered energy range. In particular, in the energy 
range relevant for experiments like P2, QWeak, 
SoLID~\cite{JeffersonLabSoLID:2022iod}, or even the 
EIC~\cite{AbdulKhalek:2021gbh}, the heavy particles 
$W^\pm$, Higgs and the top-quark do not play a role 
as dynamic degrees of freedom 
and the RGE should be defined in an effective theory 
without these particles. The solution of the RGE then 
requires matching conditions which relate the weak mixing 
angle in the different effective theories. In particular, 
$\sin^2\hat{\theta}_W(\mu)$ at an energy scale above 
the $W^\pm$-boson mass, $\mu > M_W$, describing 
the running in the full SM, must be related to 
$\sin^2\hat{\theta}_W(\mu)$ at scales $\mu < M_W$, i.e.\ 
in the framework of a low-energy effective theory. 

It is not obvious how to remove the $W^{\pm}$-bosons 
from the SM and remain with a consistent low-energy 
theory without breaking SU(2) symmetry. In 
Ref.~\cite{Hall:1980kf} the decoupling of heavy gauge bosons 
which receive their mass from spontaneous symmetry 
breaking of a grand unified theory was studied. In this 
approach it was assumed that an unbroken gauge theory 
remained below the mass scale of the symmetry breaking. 
The matching of the weak mixing angle crossing the $W$ 
threshold used in Ref.~\cite{Erler:2004in} was based on 
this result. However, just removing the $W$-boson from 
the SM, but keeping the $Z$-boson, does not correspond 
to the situation underlying the calculation in 
Ref.~\cite{Hall:1980kf}. 

We re-consider in this paper the matching of the weak 
mixing angle and suggest to base the approach on a well-defined 
low-energy effective theory which contains, apart from the 
electromagnetic interaction described by QED, 4-fermion 
operators for the parity-violating weak neutral-current 
interaction. Such an approach is well-known as 
Low-Energy Effective Field Theory, LEFT. The LEFT can be 
matched to the SM at low energies. This has been done in a 
general setting in Ref.~\cite{Dekens:2019ept}, but that 
paper does not contain the explicit result needed for the 
matching of the weak mixing angle. In the LEFT, the $Z$-boson 
does not exist; consequently the weak mixing angle can be 
introduced only as a phenomenological parameter and its 
relation to the corresponding SM parameter has to be 
determined by deriving suitable scattering matrix elements. 
To obtain the required relations, we calculate the 
low-energy coupling constants which were introduced 
long ago by Marciano and Sirlin \cite{Marciano:1982mm} 
to describe the parity-violating SM effects in atomic 
parity violation. The matching of the weak mixing angle 
in the SM to the corresponding neutral-current coupling 
in the LEFT has, to our knowledge, not been 
discussed so far. We fill this gap and correct the 
previously used prescription of Ref.~\cite{Erler:2004in}.

We remark that the running weak mixing angle, 
$\sin^2\hat{\theta}_W(\mu)$, is not an observable 
quantity by itself. It must be determined from 
observable transition or scattering amplitudes 
for which a complete calculation of higher-order 
corrections up to the required precision is needed 
(see for example Ref.~\cite{Erler:2013xha}). Therefore 
any prescription, even an ad-hoc one, could be used 
for it, provided it is then consistently used in the 
complete calculation. However, a consistent framework 
with a well-defined low-energy effective theory should 
be preferred. First, it will then be possible to obtain 
predictions for observables in a simplified approach, 
where some Feynman diagrams can be omitted or appear 
simpler to calculate, as for example 3-point vertex 
corrections generated from box graphs when the propagator 
of a heavy boson can be contracted to a 4-fermion vertex. 
This could in particular be a viable approach for the 
calculation of some two-loop corrections like the 
$\gamma\gamma Z$ or $\gamma WW$ box graphs. In any 
case, such simplified approaches will be very welcome 
as they provide cross-checks of the more complicated 
complete calculations. Second, a well-defined consistent 
framework will help to analyse results from searches  
for new physics, beyond the SM, when many measurements 
at different energy scales have to be combined. 

\bigskip

The layout of this paper is as follows: First we will 
summarize the results of a one-loop calculation of 
parity-violating scattering of two fermions in the SM, 
i.e.\ describe the low-energy constants $g_{VA}^{f_1f_2}$ 
and $g_{AV}^{f_1f_2}$ which had been introduced with the 
notation $C_1$ and $C_2$ by Marciano and Sirlin in 
Ref.~\cite{Marciano:1982mm}. Then we describe the naive 
approach for the matching of the weak mixing angle 
and, in section \ref{sec:Effective_theory}, 
the calculation of one-loop corrections in the LEFT, 
from which the definition of the weak mixing angle and 
its matching conditions can be derived. We conclude in 
section \ref{sec:Discussion}.

\section{Low-Energy Couplings in the Standard Model}
\label{sec:SM_result}

In the SM, the neutral-current parity-violating interaction 
is induced by $Z$-boson exchange. At low energies, the resulting 
dependence on the $Z$-boson mass, $M_Z$, can be removed by the 
one-loop SM relation with the Fermi constant usually given in 
the on-shell scheme (OS): 
\begin{equation}
\label{eq:fermi_constant_1loop}
G_F = \frac{\pi\alpha}{\sqrt{2}M_Z^2}
      \frac{1}{c_W^2s_W^2} \frac{1}{1-\Delta r_{\rm OS}} 
\, ,
\end{equation}
where $\Delta r_{\rm OS}$ accounts for the higher-order corrections 
of the muon decay in the SM. $\alpha$ is the fine structure 
constant and $s_W^2$, $c_W^2 = 1 - s_W^2 = M_W^2/M_Z^2$ are 
used for sine and cosine of the weak mixing angle, $s_W = 
\sin\theta_W$, defined in the on-shell scheme. The OS result 
for the correction $\Delta r_{\rm OS}$ 
reads~\cite{Sirlin:1980nh,Hollik:1988ii}
\begin{equation}
\Delta r_{\rm OS} 
= 
\mbox{Re}\left(
\frac{\left. \Sigma^{WW}_T(0)\right|_{\rm OS}}{M_W^2} 
\right) 
+ \frac{\alpha}{4\pi s_W^2}
  \left(6 + \frac{7-4s_W^2}{2s_W^2}\ln c_W^2 \right),
\end{equation}
where $\Sigma_T^{WW}(0)|_{\rm OS}$ is the transverse part 
of the OS renormalized self-energy of the $W$-boson evaluated 
at zero momentum transfer. 

A corresponding relation for the Fermi constant in terms 
of the $\MSB$-renormalized weak mixing angle was introduced 
in Refs.~\cite{Sirlin:1989uf,Degrassi:1990tu}. To keep the 
notation simple, we write $\MSB$-renormalized quantities 
without an additional hat; in particular for the sine and 
cosine of the weak mixing angle in the $\MSB$ scheme we use 
the abbreviations $s^2 = \sin^2\hat\theta_W(\mu)$ and 
$c^2 = 1 - s^2$. Then we have 
\begin{equation}
\label{eq:fermi_constant_1loop_MSbar}
G_F = \frac{\pi\alpha}{\sqrt{2}M_Z^2}
      \frac{1}{c^2 s^2} \frac{1}{1-\Delta r} 
\, . 
\end{equation}
The corresponding correction $\Delta r$ in the $\MSB$ 
scheme (see Refs.~\cite{Sirlin:1989uf,Degrassi:1990tu}),  
keeping the dependence on the $\MSB$ renormalization scale 
$\mu$, reads  
\begin{equation}
\label{eq:delta_r_SM}
\begin{multlined}[0.61\displaywidth]
\Delta r 
= 
\mbox{Re}\left(
     \frac{\Sigma^{WW}_T(0)}{\hat{M}_W^2}
   - \frac{\Sigma^{ZZ}_T\bigl(\hat{M}_Z^2\bigr)}{\hat{M}_Z^2}
   \right)
\\
- 2\frac{\delta e}{e}
+ \frac{\alpha}{4\pi s^2}
  \left[
        6
      + \frac{7 - 12 s^2}{2 s^2} \ln c^2
      + 4\ln\frac{\mu^2}{M_Z^2}
  \right].
\end{multlined}
\end{equation}
where $\Sigma^{WW}_T$ and $\Sigma^{ZZ}_T$ are the $\MSB$ 
renormalized transverse parts of the $W$ and $Z$ self 
energies, $e$ and $\delta e$ are the electric 
charge and its counterterm in the $\MSB$ scheme 
and $\hat{M}_W$, $\hat{M}_Z$ the boson masses in the $\MSB$ 
scheme. This allows us to write the relation for $G_F$ in 
terms of the $Z$-coupling $g_Z = e / sc$, and by also 
expressing the OS mass in terms of the $\MSB$-mass via 
$M_Z^2 = \hat{M}_Z^2-\mbox{Re}\,\Sigma^{ZZ}_T\bigl(M_Z^2\bigr)$, 
one can write~\cite{Degrassi:1990tu}\footnote{ 
  Only here we need to distinguish between OS ($M_W$, $M_Z$) 
  and $\MSB$ masses ($\hat{M}_W$, $\hat{M}_Z$). Below, 
  when we calculate corrections at one-loop precision, we 
  can use again the notation $M_W$ and $M_Z$ without hats 
  since the difference will be of higher order. 
  } 
\begin{equation}
\label{eq:relation_GF_fundamental_MSB}
\begin{aligned}
G_F 
= 
\frac{g_Z^2}{4\sqrt{2}\hat{M}_Z^2}
   \Bigg\{
          1 + \frac{\Sigma^{WW}_T(0)}{\hat{M}_W^2}
            + \frac{\alpha}{4\pi s^2}
              \bigg[6 + 
                    \frac{7 - 12 s^2}{2 s^2}
                    \ln c^2 
                    + 4\ln\frac{\mu^2}{M_Z^2} 
              \bigg]
            + \mathscr{O}\bigl(\alpha^2\bigr)
   \Bigg\}.
\end{aligned}
\end{equation}

\bigskip 

We consider scattering of two fermions, $f_1$, $f_2$, at low 
energy and low momentum transfer. Since we are interested 
in the weak mixing angle, we take into account only the 
parity-violating contributions. The matrix element can be 
written in the form 
\begin{equation}
\label{eq:PV_interaction_parametrization}
\mathscr{M}_\textrm{PV}\bigl(q^2\bigr)
= 
i\frac{G_F}{\sqrt{2}}
\left[
      \tilde{g}_{VA}^{f_1f_2}\bigl(q^2\bigr)
      \overline{u}_1 \gamma_\mu u_1\:
      \overline{u}_2 \gamma^\mu\gamma_5 u_2
    + \tilde{g}_{AV}^{f_1f_2}\bigl(q^2\bigr)
      \overline{u}_1 \gamma_\mu\gamma_5 u_1\:
      \overline{u}_2 \gamma^\mu u_2
\right]
+ ...\, ,
\end{equation} 
where $u_1$ and $u_2$ are the spinors of the two fermions 
$f_1$ and $f_2$. As indicated, there may be additional 
momentum-transfer dependent terms from scalar and 
pseudo-scalar couplings; these do not contribute at zero 
momentum transfer. At zero momentum transfer, this 
parametrization corresponds to the effective contact 
interaction constants $C_{1q} = \tilde{g}_{VA}^{eq}(0)$, 
$C_{2q} = \tilde{g}_{AV}^{eq}(0)$, $q=u,\, d$, introduced 
in Ref.~\cite{Marciano:1982mm}.
Due to the symmetry under the exchange $f_1\leftrightarrow f_2$, 
we can restrict the discussion to the case 
$\tilde{g}_{VA}^{f_1f_2}\bigl(q^2\bigr)$, i.e.\ 
$g_{AV}^{f_1f_2} = g_{VA}^{f_2f_1} 
= \tilde{g}_{AV}^{f_1f_2}\bigl(0\bigr)$. 

We summarize the results in the following, given in $\MSB$ 
renormalization. The low-energy coupling parameters 
$g_{VA}^{f_1f_2}$ including one-loop corrections can be 
written in the following way: 
\begin{equation}
\begin{aligned}
\label{eq:C_1f1f2_MarcianoSirlin_similar}
g_{VA}^{f_1f_2}
&= 
- 2 a_{f_2}\rho\left[T^3_{f_1} - 2Q_{f_1}\kappa s^2\right]
+ \Box
+ 4a_{f_2}v_{f_1}\frac{\alpha}{4\pi}
  \left(Q_{f_1}^2 + Q_{f_2}^2\right) 
\\
& + \frac{\alpha}{9\pi}Q_{f_1}a_{f_2}
      Q_{f_2}v_{f_2}\left(1 - 6\ln\frac{m_{f_2}^2}{M_Z^2}\right)
      + \frac{\alpha}{18\pi} Q_{f_1} Q_{\tilde{f}_2} 
      \left(1 - 6\ln\frac{m_{\tilde{f}_2}^2}{M_W^2}\right) .
\end{aligned}
\end{equation} 
We have used the usual vector and axial-vector coupling 
constants of the $Z$-boson, $v_f = T_3^f - 2 s^2 Q_f$ and 
$a_f = T_3^f$, written in terms of the third component of the 
isospin $T_3^f$ and the charge $Q_f$ of a fermion. The last 
term in the second line is coming from vertex corrections 
with a $W$ boson in the loop and is therefore proportional to 
the charge of the isospin partner of fermion $f_2$,  
$Q_{\tilde{f}_2} = Q_{f_2} - 2 T_{f_2}^3$. The universal, 
i.e.\ fermion-type independent parts of the corrections 
are absorbed into 
\begin{align}
\label{eq:rho_explicit_Marciano_Sirlin}
\rho &= 
1 + \frac{\alpha}{4\pi}
    \left\{
           \frac{3}{4s^4}\ln c^2
           - \frac{7}{4s^2}
           + \frac{3}{4s^2}\frac{m_t^2}{M_W^2}
           + \frac{3\xi}{4s^2}
               \left[
                     \frac{\ln\frac{c^2}{\xi}}{c^2-\xi}
                   + \frac{\ln\xi}{c^2(1-\xi)}
               \right]
    \right\}, 
\\
\label{eq:kappa_explicit_Marciano_Sirlin}
\kappa &= 
1 - \frac{\alpha}{2\pi s^2}
        \left\{
        \left(\frac{7}{2}c^2 
        + \frac{1}{12}\right)\ln\frac{\mu^2}{M_W^2}
        + \frac{7}{9} - \frac{s^2}{3}
        - \frac{1}{3}\sum_f N_f^c Q_f v_f \ln\frac{\mu^2}{m_f^2}
        \right\}.
\end{align}
This agrees with the results of Ref.~\cite{Marciano:1982mm}, 
but is more general since we keep the $\mu$-dependence while  
$\mu = M_W$ was chosen in Ref.~\cite{Marciano:1982mm}. In 
addition, the term proportional to $Q_{f_1}^2$ at the end of 
the first line of Eq.~(\ref{eq:C_1f1f2_MarcianoSirlin_similar}) 
was omitted in Ref.~\cite{Marciano:1982mm}. The $\rho$-parameter 
includes contributions from top-quark loops (mass $m_t$) and 
loops with a Higgs boson. The latter are conveniently written 
in terms of the parameter $\xi = \hat{M}_H^2 / \hat{M}_Z^2$ 
where $\hat{M}_H$ is the $\MSB$ Higgs mass. In the sum over 
all fermions $f$ we have made a color factor $N_f^c$ explicit, 
$N_f^c=3$ for the case of quarks contributing in the loops, 
otherwise $N_f^c=1$. Process-dependent terms from photonic 
corrections at the fermion vertices, proportional to squares 
or products of two charge factors, are written separately in 
Eq.~(\ref{eq:C_1f1f2_MarcianoSirlin_similar}), as well as the 
contribution from box diagrams, denoted by~$\Box$. These 
latter correction terms, obtained above in a free-quark 
calculation, will have to be combined with appropriate 
form factors to take into account effects from the hadronic 
structure. 

Finally, we note that 
Eq.~\eqref{eq:kappa_explicit_Marciano_Sirlin} can be 
used to derive the $\MSB$ running of the weak mixing angle. 
The corresponding $\beta$-function is obtained by differentiating  
with respect to $\mu^2$, 
\begin{equation}
\label{eq:s2_beta_function_explicit}
\mu^2\frac{\textrm{d}}{\textrm{d}\mu^2} 
s^2\kappa\bigl(\mu^2\bigr)
= \frac{\alpha}{\pi}
  \left(
      \frac{1}{6}\sum_f N_f^c Q_f v_f
    - \frac{43}{24}
    + \frac{7}{4} s^2
  \right).
\end{equation}
This form describes the running weak mixing angle at scales 
above the $W$-boson mass and above all fermions included 
in the sum.

\section{Matching -- General Idea and a First Attempt}
\label{sec:Matching}

What is, in the modern literature, called 'matching' was 
considered first in the context of grand unified theories 
to identify the effect of new heavy particles on the 
parameters of the SM. The idea can be sketched in the 
following way: A heavy particle of mass $M$ in a loop 
diagram contributing the term $\Sigma_h(q^2)$ to the self 
energy $\Sigma(q^2)$ of a gauge boson would lead, in the 
limit of small momentum transfer, $M^2 \gg q^2 \to 0 $, 
to a modification of the boson propagator which can be 
factorized:  
\begin{equation}
-ig_{\mu\nu} \frac{1}{q^2 + \Sigma(q^2)} 
= 
-ig_{\mu\nu} \frac{1}{q^2 + \Sigma_h(q^2) + \Sigma_r(q^2)} 
\simeq 
-ig_{\mu\nu} 
\frac{1}{1 + l_h} \, \frac{1}{q^2 + \Sigma_r(q^2)} \, . 
\end{equation} 
Assuming that $\Sigma_h(0)=0$, this leaves only a 
renormalization factor $1/(1 + l_h)$ obtained from the 
derivative of the self energy at zero momentum, i.e.\ with  
\begin{equation}
l_h = \Sigma^\prime_h(0) 
\end{equation} 
and a remainder $\Sigma_r(q^2)$ which contains only the 
low-mass particles. The heavy particle thus can be removed 
and predictions for phenomenology will not change if a 
corresponding renormalization of the remaining gauge field 
$A_\mu \rightarrow A_\mu^{\prime} = \sqrt{1 + l_h} A_\mu$ 
is performed 
which in turn requires a redefinition of the gauge coupling 
$g$ to keep the canonical normalization of the gauge kinetic 
part of the Lagrangean. Correspondingly, the interaction terms 
emerging from the covariant derivative, $D_\mu = \partial_\mu 
- i g A_\mu$, are modified: 
\begin{equation}
{\cal L}_{\text int} 
= 
g A_\mu J^\mu = g^\prime A^\prime_\mu J^\mu 
\quad \rightarrow \quad 
g = g^\prime \sqrt{1 + l_h} 
\, . 
\end{equation} 
A corresponding calculation in Ref.~\cite{Hall:1980kf} was 
based on a scenario where the gauge group of a grand unified 
theory breaks down to a simple gauge group and the gauge 
fields of the broken sector can be removed by a field 
renormalization as described above. Denoting the generators 
of the broken gauge group in the adjoint representation by 
$t_h$ one finds
\begin{equation}
l_h = -\frac{g^2}{48\pi^2}\operatorname{Tr}\bigl(t_h t_h\bigr) 
\, . 
\label{eq:l_Hall1}
\end{equation}
\begin{figure}[tb]
\begin{center}
\begin{picture}(290,80)(0,0)
\put(0,0){\includegraphics[width=5cm]{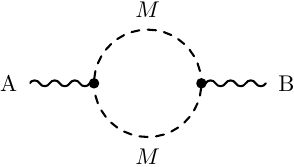}}
\put(170,10){\includegraphics[width=5cm]{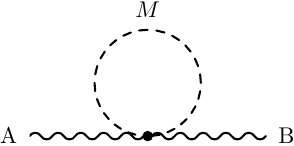}}
\end{picture}
\end{center}
\caption{ 
Topologies of one-loop diagrams with a heavy particle of 
mass $M$ contributing to the matching of the gauge coupling. 
}
\label{Fig:1loop_W_Hall}
\end{figure}

\noindent
This results from the calculation of Feynman diagrams of 
two types, shown in Fig.~\ref{Fig:1loop_W_Hall}, which 
include the heavy gauge boson as well as the corresponding 
Goldstone bosons. A naive application of the results to 
the SM, removing the charged $W^\pm$, but keeping the neutral 
component $W^3$, i.e.\ identifying 
the broken gauge group generators with those of the charged 
components of the SM gauge group SU(2)$\times$U(1), leads 
to renormalization factors $l_B$ and $l_W$ for the remaining 
gauge fields $B_\mu$ and $W^3_\mu$, 
\begin{equation}
\begin{aligned}
\label{eq:l_Hall}
l_B &= 0, 
\\
l_{W} &= -\frac{g_2^2}{24\pi^2}
\end{aligned}
\end{equation}
for the U(1) and SU(2) gauge groups, respectively. The result 
$l_B = 0$ represents the assumption that the U(1) gauge field 
does not couple to any of the fields with mass $M_W$. 
The matching of the corresponding gauge couplings $g_1$ and 
$g_2$ can then be translated into matching conditions for 
the fine structure constant and the weak mixing angle. Using 
\begin{equation}
\label{eq:s2_fundamental_definition}
s^2 = \frac{g_1^2}{g_1^2+g_2^2} \, , 
\quad \quad \quad 
4\pi\alpha = e^2 = \frac{g_1^2 g_2^2}{g_1^2+g_2^2}
\end{equation}
and the replacements $g_1 = g_1^\prime \sqrt{1+l_B} = 
g_1^\prime$, $g_2 = g_2^\prime \sqrt{1+l_W}$, one finds 
\begin{equation}
\label{eq:matching_alpha_indirectly}
\frac{1}{\alpha} 
= \frac{1}{\alpha^\prime} + \frac{1}{6\pi} 
+ \mathscr{O}(\alpha) 
\, , 
\end{equation}
and for the weak mixing angle, 
\begin{equation}
\label{eq:matching_s2_indirectly}
s^2 
= 
s^{\prime\,2} - c^2 s^2\left(l_{W}-l_B\right) 
+ \mathscr{O}\bigl(\alpha^2\bigr)
= 
s^{\prime\,2} + \frac{\alpha}{6\pi} (1 - s^{\prime\,2})
+ \mathscr{O}\bigl(\alpha^2\bigr).
\end{equation}
Equation~\eqref{eq:matching_s2_indirectly} is the matching 
condition used in 
Refs.~\cite{Erler:1998sy,Erler:2004in,Erler:2017knj} 
to determine the weak mixing angle at low energies. 
We note that in a more recent study~\cite{Martin:2018yow}, 
the matching for the fine structure constant, 
Eq.~(\ref{eq:matching_alpha_indirectly}), was confirmed 
by constructing an effective low energy theory integrating 
out the top, Higgs, $Z$ and $W$ simultaneously. Including 
the scale-dependent logarithm $L_W = \ln (\mu^2/M_W^2)$, 
the result is 
\begin{equation}
\label{eq:Martin2018_alpha_matching}
\frac{1}{\alpha} 
= 
\frac{1}{\alpha^\prime} + \frac{1}{6\pi} - \frac{7}{4\pi}L_W 
\, . 
\end{equation} 
The prescription of Eq.~(\ref{eq:matching_s2_indirectly}) 
ignores the fact that in the SM not only the charged sector 
of the gauge symmetry is broken, but spontaneous symmetry 
breaking also affects the neutral sector. This is connected 
with the fact that the Higgs field does couple to the U(1) 
gauge field, in contrast to the assumption of the above 
derivation. The derivation of matching conditions works well 
for the fine structure constant since $\alpha$ remains as the 
coupling of an unbroken U(1) symmetry; however, it does not 
work for the neutral component of the SU(2) gauge field. 
First, one cannot factorize the effect of the heavy $W$-boson, 
$(1 - \Sigma^\prime_W(0))$, from the $Z$-propagator without 
affecting also the $Z$-mass, and, secondly, the general 
formulae Eqs.~(\ref{eq:l_Hall1}, \ref{eq:l_Hall}) cannot take 
into account corrections to the $W^3$-$B$ mass mixing which 
are generated by the SM Higgs sector.

An attempt to define a low-energy theory which contains 
the photon and a massive $Z$-boson is expected to be 
incosistent: it either requires to work with a heavy boson 
which is not described by a gauge field, or one would have 
to invent a non-standard Higgs sector with other possibly 
unwanted consequences for phenomenology. Instead, we follow 
the approach to extend QED with additional neutral-current 
4-fermion operators, i.e.\ the so-called LEFT. This will 
be described in the next section. 

We note that removing a heavy fermion from the theory 
does not create the same problems at the one-loop level. 
The matching condition for the weak mixing angle when 
integrating out a single fermion $f$ is 
\begin{equation}
\label{eq:eff_contact_s2_matching_final}
s^{\prime\,2} 
= 
s^2 
+ \frac{\alpha}{6\pi} N_f^c Q_f v_f 
  \ln\frac{\mu^2}{m_f^2} 
+ \mathscr{O}\bigl(\alpha^{2}\bigr) 
\, , 
\end{equation}
which reduces simply to $s^{\prime\:2} = s^2$ at the 
fermion threshold $\mu = m_f$. An expression for the 
matching at the two-loop level, due to diagrams with 
a virtual photon inside a fermion loop, had been given 
in Ref.~\cite{Erler:2004in}.

\section{Effective Low-Energy Theory}
\label{sec:Effective_theory}

We now define the framework that allows us to describe 
parity-violating interactions at low energies in neutral-current 
processes, like elastic polarized electron scattering or 
atomic parity violation. In the SM, such interaction is 
described by the exchange of the $Z$-boson whose couplings 
to fermions are determined by the weak mixing angle. 
At low energies we consider the LEFT, consisting of QED 
extended by 4-fermion operators of dimension 6. For our 
purpose we can restrict ourselves to flavor-diagonal 
operators for the weak neutral-current with vector and 
axial-vector couplings. The ingredients that we need 
are the following: 
\begin{itemize}
\item 
The coupling constants are 
\begin{equation} 
e \rightarrow e^\prime \, , 
\quad 
G_F \rightarrow G^\prime \, , 
\quad 
s \rightarrow s^{\prime} \, , 
\end{equation}
for the electromagnetic coupling, the weak coupling replacing 
the Fermi constant, and the weak mixing angle. The $Z$-boson 
is not part of the model, but the form of its low-energy 
effective couplings are kept which amounts to include a contact 
interaction defined below. It is built from neutral-current 
operators with vector and axial-vector couplings as in the SM, 
but with 
$s^{\prime}$ instead of $s$: 
\begin{equation}
v_f^\prime = T_3^f - 2 s^{\prime\,2} Q_f \, , 
\quad 
a_f^\prime = T_3^f \, . 
\end{equation} 
We write all parameters and fields as well as the expressions 
for self energy and vertex functions of the effective 
theory with an extra prime to emphasize the difference to 
the corresponding SM quantities\footnote{
  For momentum-dependent self energies and vertex functions 
  we will put the prime as a subscript to avoid confusion 
  with a derivative. 
}. 
\item 
The $\MSB$ fermion masses, $m_f \rightarrow m_f^\prime$ are 
different at higher orders, but we do not need the 
corresponding matching relations explicitly since we are 
not interested in observables which depend on the fermion 
masses at tree level. Fermion masses appear in our calculation 
only in logarithms of one-loop corrections and the distinction 
of $m_f$ in the two theories is of higher order. 
\item 
The fields of the effective theory correspond to the SM 
photon ($A_\mu$) and fermion fields ($f_L$ and $f_R$ for 
left- and right-handed fermions), but are not identical 
with them. Matching constants $l_A$ and $l^f_{L/R}$, 
defined by 
\begin{equation} 
\label{eq:field_ren_const}
A^\prime_\mu = A_\mu \sqrt{1 + l_A}  \, , 
\quad 
f^\prime_L = f_L \sqrt{1 + l^f_L}  \, , 
\quad 
f^\prime_R = f_R \sqrt{1 + l^f_R}  \, , 
\end{equation} 
will be determined by relating the corresponding 2-point 
functions. 
\end{itemize}
The needed part of the Lagrangean is the following: 
\begin{equation}
\label{eq:effective_contact_lagrangian}
\begin{multlined}[0.58\displaywidth]
\mathscr{L}^\prime 
= 
\sum_{f} 
\left[
  \overline{f^\prime}\left(i\slashed{\partial} 
  - m_{0,f}^\prime\right)f^\prime
  - e^\prime Q_f\overline{f^\prime}\slashed{A}^\prime f^\prime
\right] 
\\
- \frac{1}{4}F^\prime_{\mu\nu}F^{\prime\mu\nu}
- \frac{1}{2\xi^\prime_A}
  \left(\partial_\mu A^{\prime\mu}\right)^2
- 4\sqrt{2}\,G^\prime 
  J_{\textrm{NC}\,\mu}^{\prime} J_\textrm{NC}^{\prime\mu} 
\, ,
\end{multlined}
\end{equation}
where the effective neutral current $J_\textrm{NC}^{\prime\mu}$ 
is defined by 
\begin{equation}
\label{eq:eff_contact_neutral_current}
J_\textrm{NC}^{\prime\mu} 
= 
\sum_f 
\left(
T^3_f\overline{f^{\prime}_L}\gamma^\mu f^{\prime}_L
- s^{\prime\,2} Q_f \overline{f^{\prime}}\gamma^\mu f^{\prime}
\right).
\end{equation} 
The Feynman rule for the 4-fermion vertex can be written in 
terms of vector and axial-vector couplings, as in the SM, 
\begin{equation}
\label{eq:eff_contact_interaction_vertex_Feynman_rule}
\raisebox{-0.47\height}{\includegraphics{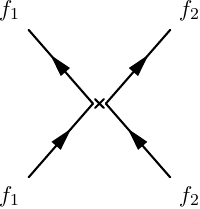}} 
\quad \corresponds \quad 
- i 2\sqrt{2}\,G^\prime \gamma_\mu 
\left(v_{f_1}^{\prime} - a_{f_1}^{\prime}\gamma_5\right)
\otimes 
\gamma^\mu\left(v_{f_2}^{\prime} - a_{f_2}^{\prime}\gamma_5\right) 
\, .
\end{equation} 
The symbol~``$\otimes$'' is used in the Feynman 
rule~\eqref{eq:eff_contact_interaction_vertex_Feynman_rule} 
to emphasize that the gamma matrices must not be contracted 
directly, but need to be wrapped between Dirac spinors associated 
with the fermion lines of the fermions $f_1$ and $f_2$.

The calculation of one-loop corrections will give rise to 
UV-divergences which have to be absorbed in properly chosen 
counterterms. There is no need in the present case to provide 
explicit results for the counterterms since we will assume 
dimensional regularization and $\MSB$ renormalization and 
simply drop the UV-divergent parts which are proportional to 
$\Delta = \frac{2}{4-D} - \gamma_\textrm{EM} + \ln 4\pi$ 
where $\gamma_\textrm{EM} = 0.577215 \ldots$ is the 
Euler--Mascheroni constant. There are, however, related 
scale-dependent logarithms, $\ln\frac{\mu^2}{M^2}$ where 
$M$ is a particle mass, which we have to carefully keep 
track of. Explicit results for the SM one-loop corrections 
can be taken for example from Ref.~\cite{Bohm:1986rj}. The 
calculations use the naive anti-commuting $\gamma_5$ scheme 
which is expected to lead to consistent results for the 
application in our case \cite{Trueman:1995ca}.

\bigskip 

The matching constant $l_A$ for the photon field can be 
obtained by relating the photonic 2-point functions in the 
two theories at zero momentum. The transverse part of the 
photon self energy is zero at $q^2=0$ due to a Ward 
identity, valid in both theories, 
$\Sigma_T^{\gamma\gamma}(q^2=0)=0$. Relating also the 
derivatives allows us to obtain $l_A$: 
\begin{equation} 
\label{eq:lA_from_Sigma_gg}
l_A = \Sigma_T^{\gamma\gamma\prime}(0) 
- \Sigma_{\prime T}^{\gamma\gamma\prime}(0)
= - \frac{\alpha}{4\pi} 
\left(3 \ln \frac{\mu^2}{M_W^2} + \frac{2}{3} \right)
\end{equation} 
which is originating from the loop diagrams with a 
$W$-boson and its corresponding Goldstone bosons, not 
present in the LEFT. 

\begin{figure}[bp]
\begin{center}
\includegraphics{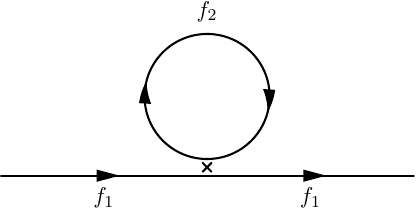} 
~~~~ 
\includegraphics{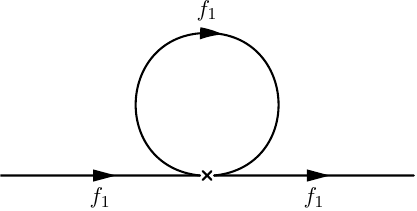}
\end{center}
\caption{
Tadpole diagrams for $\Sigma_{f}^T$ in the effective 
theory that arise from the interaction of a fermion line 
with a $Z$-boson in the Standard Model. The closed-loop 
diagram (left) vanishes. 
\label{fig:eff_contact_fermion_self_energy_tadpole}
}
\end{figure}
Calculating the one-loop fermion propagator allows us to 
determine the fermion field matching constants $l^f_{L/R}$. 
In the LEFT there is a contribution to the fermion self 
energy with a topology which is not present in the SM. 
There are two Feynman diagrams shown in 
Fig.~\ref{fig:eff_contact_fermion_self_energy_tadpole}. 
The one with a closed fermion loop (left part of 
Fig.~\ref{fig:eff_contact_fermion_self_energy_tadpole}) 
vanishes. The second one (right part of 
Fig.~\ref{fig:eff_contact_fermion_self_energy_tadpole}) 
leads to 
\begin{equation}
\Sigma_{f}^T 
= - \frac{\sqrt{2}\,G^\prime}{4\pi^2} 
\left(v_{f}^2 - a_{f}^2\right)
m_{f}^3\left(\Delta + \ln\frac{\mu^2}{m_{f}^2} 
+ \frac{1}{2}\right)
\end{equation}
which gives a contribution to the scalar part of the fermionic 
self-energy function, only. Note that here and below we omit 
primes at the couplings if they appear in correction terms 
since the difference is of higher order. The diagram in 
Figure~\ref{fig:eff_contact_fermion_self_energy_tadpole} is 
a tadpole diagram in which the loop does not depend on the 
momentum of the fermion. Therefore the corresponding self-energy 
function only depends on its mass. As a consequence, 
the loop merely induces a shift of the fermion mass $m_{f}$, 
which can be absorbed in a suitable matching constant. Since 
the distinction of the masses in the effective theory and the 
SM is irrelevant for our purpose, we ignore these tadpole 
diagrams altogether in the following. 

The result for the vector and axial-vector components of 
the fermion field matching constants, $l_V^f = 
\bigl(l_R^f + l_L^f\bigr)/2$, $l_A^f = \bigl(l_R^f - 
l_L^f\bigr)/2$ can be written as 
\begin{equation}
\label{eq:fermion_matching_l_V}
l_V^f 
= 
\Sigma_V^f\bigl(m_{f}^2\bigr)
- \Sigma_{\prime V}^f\bigl(m_{f}^2\bigr)
+ 2m_{f}^2
  \left[
  \Sigma_V^{f\prime}\bigl(m_{f}^2\bigr)
  - \Sigma_{\prime V}^{f\prime}\bigl(m_{f}^2\bigr)
  + \Sigma^{f\prime}_{S}\bigl(m_{f}^2\bigr)
  - \Sigma^{f\prime}_{\prime S}\bigl(m_{f}^2\bigr)
  \right] 
\end{equation}
and 
\begin{equation}
\label{eq:fermion_matching_l_A}
l_A^f 
= 
\Sigma_A^f\bigl(m_{f}^2\bigr)
- \Sigma_{\prime A}^f\bigl(m_{f}^2\bigr)
\, .
\end{equation} 
The axial part of the fermion self energy does in fact 
not contribute at one-loop order in the effective theory, 
$\Sigma_{\prime A}^f\bigl(m_{f}^2\bigr) = 
\mathscr{O}\bigl(\alpha^2\bigr)$,  
and the differences of the one-loop expressions in the two 
theories is due to loops with a $Z$- or $W$-boson and read 
explicitly, omitting contributions of higher-order and 
terms of order 
$\mathscr{O}\left(m_f^2 / M_{W,Z}^2\right)$: 
\begin{equation}
\label{eq:fermion_matching_l_V_ex}
l_V^f 
= 
\frac{\alpha}{4\pi}
\left\{\frac{v_f^2 + a_f^2}{4c^2s^2}
         \left(\ln\frac{\mu^2}{M_Z^2} - \frac{1}{2}\right)
       + \frac{1}{4s^2}
         \left(\ln\frac{\mu^2}{M_W^2} - \frac{1}{2}\right)
\right\}
\end{equation}
and 
\begin{equation}
\label{eq:fermion_matching_l_A_ex}
l_A^f 
= 
- \frac{\alpha}{4\pi}
\left\{\frac{v_f a_f}{2c^2s^2}
         \left(\ln\frac{\mu^2}{M_Z^2} - \frac{1}{2}\right)
       + \frac{1}{4s^2}
         \left(\ln\frac{\mu^2}{M_W^2} - \frac{1}{2}\right)
\right\} .
\end{equation}

\bigskip 

Next we verify the matching relation for the electromagnetic 
coupling. This can be obtained from relating the full 3-point 
function describing the photon-fermion coupling,  
\begin{equation}
\label{eq:matching_three_point_ff_gamma_introduction}
\overline{u}(p)\Gamma^{ff\gamma}_{\prime\mu}(p, p)u(p) 
= 
\overline{u}(p)\Gamma^{ff\gamma}_\mu(p, p)u(p)
\end{equation}
in the LEFT and the SM for some value of the momentum $p$ 
with $p^2 = m_f^2$. Scalar and pseudo-scalar contributions 
to the vertex functions do not contribute to the matching. 
The one-loop expansion of the vector part of 
Eq.~\eqref{eq:matching_three_point_ff_gamma_introduction} 
allows us to write  
\begin{equation}
\begin{aligned}
\label{eq:eff_contact_matching_condition_EM_coupling_1loop}
\frac{e^\prime}{e}
= 1
  &- \frac{\Sigma_T^{\gamma\gamma\prime}(0) 
      - \Sigma_{\prime T}^{\gamma\gamma\prime}(0)}{2}
  - \left(
          \Sigma_V^{f}\bigl(m_f^2\bigr)
        - \Sigma_{\prime V}^{f}\bigl(m_f^2\bigr)
    \right) 
\\
  &+ \frac{\Lambda^{ff\gamma}_{V}(0) 
      - \Lambda^{ff\gamma}_{\prime V}(0)}{Q_f}
   - \frac{v_f}{2csQ_f}\frac{\Sigma^{\gamma Z}_{T}(0)}{M_Z^2}
   + \mathscr{O}\bigl(\alpha^2\bigr)
\, . 
\end{aligned}
\end{equation} 
\begin{figure}[tbp]
\begin{center}
\includegraphics{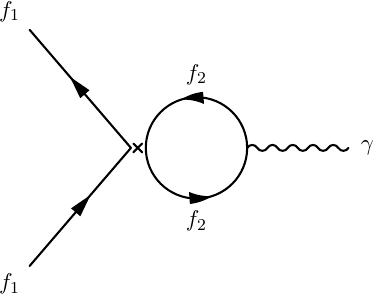}
~~~~ 
\includegraphics{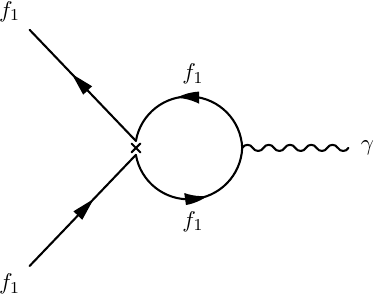}
\end{center}
\caption{
Feynman diagrams with a 4-fermion vertex contributing to 
the $ff\gamma$-vertex at one-loop order in the effective 
contact interaction model. There are two ways to contract 
the fermion lines with the 4-fermion vertex. The diagram 
at the left side arises from a $Z$-boson propagator connecting 
the external fermion and the loop. The diagram at the right 
side corresponds to the SM vertex correction with a $Z$-boson 
in the loop. 
}
\label{fig:eff_contact_vertex_correction}
\end{figure}

\noindent
The differences of the photon and fermion self energies 
have been used already above, see 
Eqs.~(\ref{eq:lA_from_Sigma_gg}) and 
(\ref{eq:fermion_matching_l_V},\ref{eq:fermion_matching_l_V_ex}).  
For the proper vertex diagrams, $\Lambda^{ff\gamma}_{\prime}$, 
we have to take into account a new contribution shown in 
Fig.~\ref{fig:eff_contact_vertex_correction} generated by 
the 4-fermion vertex in the LEFT. Using the generic 
notation $V_1$, $A_1$, $V_2$ and $A_2$ for the vector and 
axial-vector couplings at the left- and right-hand side 
of the loop, one finds 
\begin{equation}
\label{eq:self_energy_generic_closed_fermion_loop}
\begin{aligned}
\Sigma_\prime^{\bigcirc_{\!f}}\!\bigl(q^2\bigr) 
= \frac{1}{16\pi^2}\frac{4}{3}\sum_{f}N_f^c
  \bigg\{
  & 
  6A_1 A_2 m_f^2\Delta_f 
  \\
  & 
  + q^2\Big[
    A_1 A_2 - \left(A_1 A_2 + V_1 V_2\right)\Delta_f
    \Big]
  + \mathscr{O}\bigl(q^4\bigr)
  \bigg\}
\end{aligned}
\end{equation}
when expanded at zero momentum transfer. We denote this result 
for the loop integral by $\Sigma_\prime^{\bigcirc_{\!f}}$ 
since it is a two-point function although it contributes to 
the vertex correction $\Lambda^{ff\gamma}_{\prime V}$. 
The $A_1 A_2$-terms do not contribute to the Feynman diagram 
depicted in Figure~\ref{fig:eff_contact_vertex_correction} due 
to the coupling to the photon, $A_2=0$. Inserting 
$V_1 = - i \sqrt{2} G^\prime v_{f_2}$, 
$V_2 = -ieQ_{f_2}$ and omitting the UV-divergent 
parts proportional to $\Delta$ allows to write the complete 
vertex function $\Lambda_{\prime V}^{ff\gamma}$ as 
\begin{equation}
\Lambda_{\prime V}^{ff\gamma}\bigl(q^2\bigr)
= 
  \begin{aligned}[t] 
  & 
  \frac{\alpha}{4\pi}Q_{f}^3
  \left(
  \ln\frac{\mu^2}{m_f^2} + 2\ln\frac{\lambda^2}{m_{f}^2} + 2
  \right) 
  \\ &
- \frac{i\sqrt{2}\,G^\prime}{4\pi^2}\frac{e}{3} q^2 
  \Bigg\{ 
  v_{f_1} \sum_{f_2} N_{f_2}^c Q_{f_2} v_{f_2}
  \Bigl(\ln\frac{\mu^2}{m_{f_2}^2} \Bigr) 
  + 
  Q_{f_1} \frac{v_{f_1}^2 + a_{f_1}^2}{4} 
  \Bigl( \ln\frac{\mu^2}{m_{f_1}^2} - 1 
   \Bigr) 
  \Bigg\} + \mathscr{O}\bigl(q^4\bigr)
  . 
  \end{aligned}
\end{equation} 
There is an infrared divergence, as in the SM, which we 
regularize with the help of a finite, infinitesimally small 
photon mass $\lambda$. At zero momentum transfer only the 
first term from the SM photon-fermion vertex correction 
survives and the difference of the vertex corrections in 
the SM and the LEFT is only due to the loops with a $Z$- 
or a $W$-boson:  
\begin{equation}
\Lambda^{ff\gamma}_{V}(0) - \Lambda^{ff\gamma}_{\prime V}(0)
= 
\frac{\alpha}{4\pi}
  \begin{aligned}[t]
  \Bigg\{
        & Q_f\frac{v_f^2 + a_f^2}{4c^2s^2}
         \left(\ln\frac{\mu^2}{M_Z^2} - \frac{1}{2}\right) 
    \\
        &+ \frac{Q_{\tilde{f}}}{4s^2}
         \left(\ln\frac{\mu^2}{M_W^2} - \frac{1}{2}\right)
         + \frac{3T^3_f}{2s^2}
         \left(\ln\frac{\mu^2}{M_W^2} - \frac{1}{6}\right)
  \Bigg\} 
  \, .
  \end{aligned}
\end{equation} 
The second term inside the braces is due to the vertex 
correction with a $W$-boson and the isospin partner 
$\tilde{f}$ of the external fermion $f$ inside the loop and 
the last term in the second line is due to the non-Abelian 
$WW\gamma$ coupling. We also need the $\gamma Z$ mixing 
in the SM,  
\begin{equation}
\label{eq:sigma_gZ_0}
\Sigma_T^{\gamma Z}(0) 
= 
\frac{\alpha}{4\pi} 
\left(2\frac{M_W^2}{cs}\ln\frac{\mu^2}{M_W^2}\right) 
\, . 
\end{equation} 
Collecting these results in 
Eq.~\eqref{eq:eff_contact_matching_condition_EM_coupling_1loop} 
and using $Q_f - Q_{\tilde{f}} = 2 T_f^3$, we observe that 
terms from the fermion self-energies cancel against corresponding 
terms from vertex corrections. Also all terms that appear to 
depend on the isospin, $T_f^3$, partly due to the non-Abelian 
coupling of the photon to the $W$ boson in the SM, drop out. 
The remainder is determined from the photon self energy and 
the $\gamma Z$ mixing, and we find the explicit, final  
expression for the effective electromagnetic coupling constant 
in the LEFT in terms of the SM \(\MSB\)-renormalized coupling,
\begin{equation}
\frac{e^\prime}{e} 
= 1 + \frac{\alpha}{4\pi}
      \left(\frac{1}{3}
           + \frac{7}{2}\ln\frac{\mu^2}{M_W^2}
      \right), 
\end{equation}
or
\begin{equation}
\label{eq:eff_contact_EM_matching_final}
\frac{1}{\alpha} = \frac{1}{\alpha^\prime}
                       + \frac{1}{6\pi}
                       + \frac{7}{4\pi}\ln\frac{\mu^2}{M_W^2}.
\end{equation}
Equation~\eqref{eq:eff_contact_EM_matching_final} is identical 
to the matching condition found in Ref.~\cite{Martin:2018yow} 
and its scale-independent part was used in 
Refs.~\cite{Erler:2004in,Erler:1998sy,Erler:2017knj} to 
match the running fine structure constant at the $W$ threshold.

\bigskip 

To obtain a matching condition for the weak mixing angle 
we calculate the low-energy limit of the parity-violating 
contribution to fermion scattering, expressed in terms of 
the low-energy constants $g_{VA}^{f_1f_2}$ for the 
``vector--axial~vector''-part of the scattering matrix 
element. At tree-level one needs the 4-fermion vertex which 
we combine with the LSZ factors $\dLSZ_V^f$ and $\dLSZ_A^f$ 
to account for the self energy corrections at the external 
fermion legs since in the $\MSB$ scheme the residues of the 
propagators are not renormalized to unity. We label this first 
contribution with an index $(A)$. For one fermion line we have 
\begin{equation}
\begin{aligned}
&
\overline{u}\bigl(p^\prime\bigr)
\left(1 - \frac{1}{2}\dLSZ_V^f + \frac{1}{2}\dLSZ_A^f\gamma_5\right)
\gamma_\mu\left(v_f^{\prime} - a_f^{\prime} \gamma_5\right)
\left(1 - \frac{1}{2}\dLSZ_V^f - \frac{1}{2}\dLSZ_A^f\gamma_5\right)
u(p) 
\\
=\ 
& 
\overline{u}\bigl(p^\prime\bigr)
\gamma_\mu
\left(v_f^{\prime} - a_f^{\prime} \gamma_5\right)
\left[1 - \Sigma_V\bigl(m_f^2\bigr)
        - 2m_f^2\left(
                    \Sigma_V^\prime\bigl(m_f^2\bigr)
                  + \Sigma_S^\prime\bigl(m_f^2\bigr)
              \right)
\right]
u(p) \, .
\end{aligned}
\end{equation} 
Inserting the explicit result for the vector and scalar 
parts of the fermion self energy, omitting the UV divergent 
contributions, we find 
\begin{equation}
\label{eq:eff_contact_fermion_line_with_LSZ}
\overline{u}\bigl(p^\prime\bigr)
\gamma_\mu \left(v_f^{\prime} - a_f^{\prime}\gamma_5\right)
u(p)
\cdot  
\left[1 - \frac{\alpha}{4\pi}Q_f^2
          \left(\ln\frac{\mu^2}{m_f^2}
                + 2\ln\frac{\lambda^2}{m_f^2}
                + 4
          \right)
\right] 
\end{equation}
and the ``vector--axial~vector''-part of the entire tree-level 
matrix element including fermion self-energy corrections 
becomes
\begin{equation}
\begin{multlined}[0.8\displaywidth]
i \sqrt{2}G^\prime \left(v_{f_1}^{\prime}a_{f_2}^{\prime}\right)
\overline{u}\bigl(p_1^\prime\bigr)
\gamma_\mu 
u\bigl(p_1\bigr) \;
\overline{u}\bigl(p_2^\prime\bigr)\gamma^\mu\gamma_5 u\bigl(p_2\bigr)
\\ 
\cdot
\left[
  1 - \frac{\alpha}{4\pi}\sum_{f=f_1,f_2} Q_f^2
      \left(\ln\frac{\mu^2}{m_f^2}
            + 2\ln\frac{\lambda^2}{m_f^2}
            + 4
      \right)
\right].
\end{multlined}
\end{equation}
This contributes to the effective coupling constant 
$g_{VA}^{f_1f_2\prime}$ in the effective model: 
\begin{equation}
\label{eq:eff_contact_C1f1f2_A}
g_{VA}^{f_1f_2(A)\prime} 
= 
- 2 \frac{G^\prime}{G_F} 
  \, v_{f_1}^{\prime}a_{f_2}^{\prime} \, 
  \left[1 - \frac{\alpha}{4\pi}\sum_{f=f_1,f_2} Q_f^2
            \left(\ln\frac{\mu^2}{m_f^2} 
                  + 2\ln\frac{\lambda^2}{m_f^2}
                  + 4
            \right)
\right].
\end{equation} 
$G_F$ appears in the denominator because the dimensionless 
couplings $g_{VA}^{f_1f_2\prime}$ are defined with respect to 
the SM result, Eq.~(\ref{eq:PV_interaction_parametrization}). 

\begin{figure}[tbp]
\newcommand{\subfigwidth}{.43}
\hfill
\begin{minipage}{\subfigwidth\textwidth}
\centering
\includegraphics{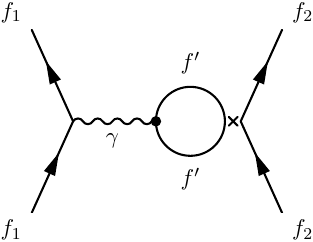}
\subcaption{}
\label{fig:contact_interaction_vertex_correction_A_PV}
\end{minipage}
\hfill
\begin{minipage}{\subfigwidth\textwidth}
\centering
\includegraphics{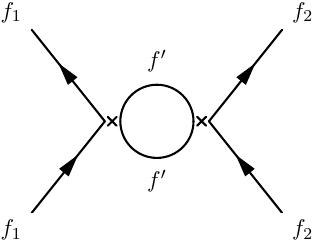}
\subcaption{}
\label{fig:contact_interaction_double_contact_correction}
\end{minipage}
\hfill\hfill
\caption{
Diagrams with fermion loop insertions describing the 
contributions ($B$) and ($C$) to the parity violating 
interaction in the effective interaction model. 
}
\label{fig:eff_contact_A_PV_fermion_loop_diagrams}
\end{figure}

There are two one-loop diagrams with a closed fermion loop 
contributing to the parity-violating interaction, shown in 
Figure~\ref{fig:eff_contact_A_PV_fermion_loop_diagrams}.
In diagram~\ref{fig:contact_interaction_vertex_correction_A_PV}, 
the photon propagator accounts for a factor $q^{-2}$ which 
induces a contribution in terms of the derivative of 
$\Sigma^{\bigcirc_{\!f}}$, 
Eq.~(\ref{eq:self_energy_generic_closed_fermion_loop}), at 
zero momentum transfer. Leaving out terms that do not belong 
to the ``vector--axial~vector''-part of the matrix element, 
the Feynman diagram in 
Figure~\ref{fig:contact_interaction_vertex_correction_A_PV} 
translates to
\begin{equation}
\begin{multlined}
\overline{u}\bigl(p_1^\prime\bigr)
\left(-ieQ_{f_1}\gamma_\mu\right)u\bigl(p_1\bigr)
\frac{-i}{q^2}
\left(-i\Sigma^{\bigcirc_{\!f}}\!\bigl(q^2\bigr)\right)
\overline{u}\bigl(p_2^\prime\bigr)
\left(i\sqrt{2}G^\prime a_{f_2}\gamma^\mu\gamma_5\right)
u\bigl(p_2\bigr) 
\\
= 
- i \sqrt{2}G^\prime \frac{\alpha}{4\pi}Q_{f_1}a_{f_2}
  \frac{4}{3}\sum_{f^\prime}N_{f^\prime}^c 
  \left(Q_{f^\prime}v_{f^\prime}
    \ln\frac{\mu^2}{m_{f^\prime}^2} \right)
\cdot 
\overline{u}\bigl(p_1^\prime\bigr)\gamma_\mu u\bigl(p_1\bigr)\;
\overline{u}\bigl(p_2^\prime\bigr)\gamma_\mu\gamma_5 
u\bigl(p_2\bigr),
\end{multlined}
\end{equation}
where the generic vector and axial vector couplings in 
Eq.~\eqref{eq:self_energy_generic_closed_fermion_loop} 
were replaced by $A_1 = 0$, $V_1 = -ieQ_{f^\prime}$ and 
$V_2 = -i \sqrt{2} G^\prime v_{f^\prime}$.
This yields the following contribution, part $(B)$, to the 
low-energy coupling: 
\begin{equation}
\label{eq:eff_contact_C1f1f2_B}
g_{VA}^{f_1f_2(B)\prime} 
= 
2 \frac{G^\prime}{G_F}
\frac{\alpha}{4\pi}Q_{f_1}a_{f_2}
\frac{4}{3}
\sum_{f^\prime}
N_{f^\prime}^c Q_{f^\prime}v_{f^\prime} 
\ln\frac{\mu^2}{m_{f^\prime}^2} 
\, . 
\end{equation}

The second diagram with a fermion loop, depicted in 
Figure~\ref{fig:contact_interaction_double_contact_correction} 
can be understood as the low energy limit of the 
$Z$-boson's self-energy diagram in the Standard Model.
The ``vector--axial~vector''-part of its matrix element reads
\begin{equation}
\label{eq:eff_contact_double_contact_C_contribution}
\begin{multlined}
\overline{u}\bigl(p_1^\prime\bigr)
\left(-i\sqrt{2}G^\prime v_{f_1}\right)
\gamma_\mu 
u\bigl(p_1\bigr)
\left(-i\Sigma_T^{\bigcirc_{\!f}}\!\bigl(q^2\right)\bigr)
\overline{u}\bigl(p_2^\prime\bigr) 
\left(i\sqrt{2}G^\prime a_{f_2}\right)
\gamma^\mu\gamma_5 
u\bigl(p_2\bigr) 
\\
= 
-i 
\overline{u}\bigl(p_1^\prime\bigr)\gamma_\mu u\bigl(p_1\bigr)
\overline{u}\bigl(p_2^\prime\bigr)\gamma^\mu\gamma_5 u\bigl(p_2\bigr)
\cdot 
2\left(G^\prime\right)^2 
\frac{v_{f_1}a_{f_2}}{16\pi^2}\sum_{f^\prime} 
N_{f^\prime}^c m_{f^\prime}^2 
\ln\frac{\mu^2}{m_{f^\prime}^2} \, ,
\end{multlined}
\end{equation}
where the generic axial vector coupling constants were replaced 
by $A_1 A_2 = a_{f^\prime}^2 = \frac{1}{4}$. The corresponding 
contribution to the low-energy coupling, part $(C)$, is 
\begin{equation}
\label{eq:eff_contact_C1f1f2_C}
g_{VA}^{f_1f_2(C)\prime} 
= 
- \frac{4\sqrt{2}\left(G^\prime\right)^2}{G_F} 
\frac{v_{f_1}a_{f_2}}{16\pi^2}\sum_{f^\prime} 
N_{f^\prime}^c m_{f^\prime}^2 
\ln\frac{\mu^2}{m_{f^\prime}^2} \, .
\end{equation}

\begin{figure}[t]
\centering
\includegraphics{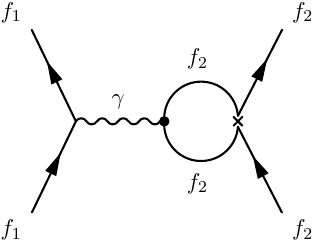}
\caption{
Diagram describing the charge radius correction, contribution 
($D$), to the parity violating low-energy coupling 
$g_{VA}^{f_1f_2}$. 
}
\label{fig:eff_contact_A_PV_charge_radius}
\end{figure}
\begin{figure}[h]
\centering
\begin{subfigure}[t]{0.49\textwidth}
\centering
\includegraphics{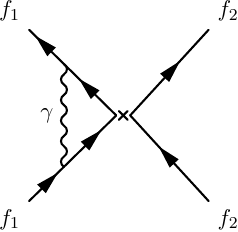}
\hfill
\subcaption{}
\label{fig:contact_interaction_photonic_correction_left}
\end{subfigure}
\hfill
\begin{subfigure}[t]{0.49\textwidth}
\centering
\includegraphics{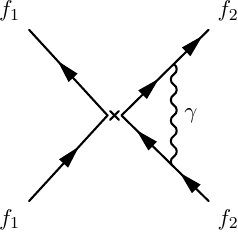}
\subcaption{}
\label{fig:contact_interaction_photonic_correction_right}
\end{subfigure}
\caption{
Photonic corrections to the parity violating interaction in 
the contact interaction model. The diagrams with a photon 
connecting the fermions $f_1$ and $f_2$ are process 
dependent remnants of the $\gamma Z$-box graphs in the 
Standard Model and will not be discussed. 
}
\label{fig:eff_contact_A_PV_photonic_corrections}
\end{figure}

The diagram shown in 
Figure~\ref{fig:eff_contact_A_PV_charge_radius} corresponds 
to what is known as charge-radius correction in the SM. There 
it is described by a one-loop vertex correction with an 
internal $Z$-boson connecting the external fermion lines at 
the vertex. This diagram contributes to $g_{VA}^{f_1f_2\prime}$, 
while a similar diagram (not shown) with the loop connected 
to fermion $f_1$ would contribute to the $AV$ coupling, 
i.e.~$g_{AV}^{f_1f_2\prime}$. We find 
\begin{equation}
\label{eq:eff_contact_C1f1f2_D}
g_{VA}^{f_1f_2(D)\prime} 
= 
2 \frac{G^\prime}{G_F}
\frac{\alpha}{4\pi}Q_{f_1}a_{f_2}
\frac{4}{3}
Q_{f_2} v_{f_2} 
\biggl(\ln\frac{\mu^2}{m_{f_2}^2} - 1 \biggr)
\, . 
\end{equation}

Finally, two one-loop diagrams are obtained by adding 
virtual photon lines to the tree-level diagram as shown in 
Figure~\ref{fig:eff_contact_A_PV_photonic_corrections}.
These and two other, similar diagrams with a photon connecting 
the two different fermions are process dependent. These other 
diagrams are linked to the $\gamma Z$-box graphs in the 
Standard Model and will be added separately. It is sufficient 
to calculate only one of the associated matrix elements of 
Figure~\ref{fig:eff_contact_A_PV_photonic_corrections}, as 
the diagrams are symmetric under the exchange of the external 
fermions. The matrix element of 
diagram~\ref{fig:contact_interaction_photonic_correction_left} 
is identical to the corresponding expression of the photonic 
vertex correction in a Standard Model calculation. This 
allows us to copy known results from above. Keeping 
``vector--axial~vector''-terms only, one finds
\begin{equation}
\label{eq:eff_contact_photonic_correction_left_VA}
i 
\overline{u}\bigl(p_1^\prime\bigr)\gamma_\mu u\bigl(p_1\bigr) 
\;
\overline{u}\bigl(p_2^\prime\bigr)
\gamma^\mu\gamma_5u\bigl(p_2\bigr)
\cdot
\sqrt{2}G^\prime v_{f_1}a_{f_2}
\frac{\alpha}{4\pi}Q_{f_1}^2
\left[
\ln\frac{\mu^2}{m_{f_1}^2} + 2\ln\frac{\lambda^2}{m_{f_1}^2} + 2
\right]
\end{equation}
for the contribution of 
diagram~\ref{fig:contact_interaction_photonic_correction_left} 
to the parity-violating interaction. The expression for 
diagram~\ref{fig:contact_interaction_photonic_correction_right} 
can be obtained from 
Eq.~\eqref{eq:eff_contact_photonic_correction_left_VA} 
by replacing the subscript $f_1$ inside the brackets with 
$f_2$. We combine these two contributions in part ($E$) of 
the low-energy coupling, $C_{1f_1}^{f_2(E)\prime}$:  
\begin{equation}
\label{eq:eff_contact_C1f1f2_E}
\begin{aligned}
g_{VA}^{f_1f_2(E)\prime} 
&= 
- \frac{2 G^\prime}{G_F} v_{f_1}a_{f_2}\frac{\alpha}{4\pi}
\left(V\bigl(f_1\bigr) + V\bigl(f_2\bigr) \right), 
\\
V(f) 
&= 
Q_f^2\left(\ln\frac{\mu^2}{m_{f}^2}
          + 2\ln\frac{\lambda^2}{m_f^2}
          + 2
     \right).
\end{aligned}
\end{equation}
Hence, the sum of all contributions to the effective low 
energy coupling in the contact interaction model,  
Eqs.~\eqref{eq:eff_contact_C1f1f2_A}, 
\eqref{eq:eff_contact_C1f1f2_B}, 
\eqref{eq:eff_contact_C1f1f2_C},  
\eqref{eq:eff_contact_C1f1f2_D} 
and~\eqref{eq:eff_contact_C1f1f2_E} reads 
\begin{equation}
\label{eq:eff_contact_C1_final}
\begin{aligned}
g_{VA}^{f_1f_2\prime} 
= 
& - 2 a_{f_2}\rho^\prime
    \left(T^3_{f_1}
          - 2Q_{f_1}\kappa^\prime s^{\prime\,2}
    \right) 
\\
& + \frac{G^\prime}{G_F} 4 v_{f_1}a_{f_2}
    \frac{\alpha}{4 \pi}
    \left(Q_{f_1}^2 + Q_{f_2}^2\right)
+ \Box^\prime 
+ \mathscr{O}\bigl(\alpha^{2}\bigr) 
\\ 
&
  + \frac{G^\prime}{G_F}
  \frac{\alpha}{9\pi} Q_{f_1}a_{f_2} Q_{f_2} v_{f_2} 
  \biggl(-6 -6 \ln\frac{m_{f_2}^2}{\mu^2} \biggr)
\, , 
\end{aligned}
\end{equation}
where we have added $\Box^\prime$ to denote the $\gamma Z$-box 
remnants and where the process independent parameters 
$\rho^\prime$ and $\kappa^\prime$ are defined as
\begin{subequations}
\label{eq:eff_contact_kappa_rho_final}
\begin{align}
\rho^\prime 
&= 
\frac{G^\prime}{G_F} 
\left(1 - \frac{G^\prime}{4\sqrt{2} \pi^2}
\sum_f N_f^c m_f^2 \ln\frac{\mu^2}{m_{f}^2} \right) \, , 
\\
\kappa^\prime 
&= 
1 + \frac{\alpha}{6\pi s^2} \sum_f N_f^c Q_f v_f 
\ln\frac{\mu^2}{m_{f}^2} \, .
\end{align}
\end{subequations}
Comparing this result with the SM expression in 
Eq.~\eqref{eq:C_1f1f2_MarcianoSirlin_similar} we observe 
that photonic corrections which are assigned to a single 
external fermion and proportional to the square of the 
charge are identical in the two theories (apart from the 
normalization factor $G^\prime /G_F$). The other 
process-dependent correction terms (the second line in 
Eq.~\eqref{eq:C_1f1f2_MarcianoSirlin_similar}) are 
only partly recovered: the charge-radius correction from 
the vertex with a $Z$-boson in the loop appears with the 
renormalization scale instead of $M_Z$ in the logarithm 
and with a different constant. The charge-radius correction 
due to the $W$-boson loop in the SM is not found in our 
calculation since we did not include a charged-current 
interaction term in the model. 

\bigskip 

The parameters of the LEFT, $G^\prime$ and $s^\prime$ can 
now be determined by equating the process-independent terms 
of the low-energy couplings $g_{VA}^{f_1f_2\prime}$ and 
$g_{VA}^{f_1f_2}$, i.e.\ 
\begin{subequations}
\label{eq:s2_rho_matching_generic}
\begin{align}
  \rho^\prime &= \rho, 
  \\
  \kappa^\prime s^{\prime\,2} &= \kappa s^2.
\end{align}
\end{subequations}
The expression for the effective coupling strength 
$G^\prime$ is obtained from the equation for the 
$\rho$-parameters. With the Standard Model expression 
given in Eq.~\eqref{eq:rho_explicit_Marciano_Sirlin}, 
one finds
\begin{equation}
\label{eq:eff_contact_K_final}
G^\prime 
= 
G_F \rho
+ 
\frac{G_F\alpha}{8\pi c^2s^2}\sum_f N_f^c \frac{m_f^2}{M_Z^2} 
\ln\frac{\mu^2}{m_{f}^2}
+ \mathscr{O}\bigl(\alpha^{2}\bigr) ,
\end{equation}
where one factor of the Fermi constant in the second term 
was replaced with its definition in terms of the electromagnetic 
coupling constant $\alpha$. The correction in 
Eq.~\eqref{eq:eff_contact_K_final} can be traced back to the 
Feynman diagram in 
Fig.~\ref{fig:eff_contact_A_PV_fermion_loop_diagrams}b, 
i.e.\ the low-energy limit of the $Z$ self energy. 
Finally, combining the SM result 
\eqref{eq:kappa_explicit_Marciano_Sirlin} with  
\eqref{eq:eff_contact_kappa_rho_final} yields
\begin{align}
\label{eq:effective_s2_final-4F}
s^{\prime\,2} 
&= 
s^2 
- \frac{\alpha}{2\pi}
  \left\{
         \left(
               \frac{7}{2}c^2 + \frac{1}{12}
         \right) \ln\frac{\mu^2}{M_W^2}
       + \frac{7}{9} - \frac{s^2}{3}
  \right\} . 
\end{align}
These terms are identical to the corresponding ones in the 
Standard Model, all originating from loop diagrams with a 
$W$-boson. The numerical difference between 
Eq.~\eqref{eq:effective_s2_final-4F} 
and~\eqref{eq:matching_s2_indirectly} for $\mu = M_W$ is 
\begin{equation}
\label{eq:effective_s2_final_diff_JE}
\Delta s^{\prime\,2}
= 
\frac{\alpha}{\pi} \cdot \frac{2}{9}
	  + \mathscr{O}\bigl(\alpha^2\bigr)
	\approx 5.2 \cdot 10^{-4} . 
\end{equation}
This is a small shift compared to the numerical value 
of the weak mixing angle ($\Delta s^{\prime\,2} / s^2 
\approx 0.2{\%}$). It has been noticed already in 
Ref.~\cite{Erler:2022ckm} that a consistent framework 
requires to include this term. There it was found that 
the term $2\alpha/(9\pi)$ is needed to bring the 1-loop 
effective weak mixing angle calculated in 
Ref.~\cite{Czarnecki:1995fw} for polarized 
Moller scattering in agreement with the running weak mixing 
angle evolved down to zero momentum transfer when the RGE 
running is based on the prescription of 
Ref.~\cite{Erler:2004in,Erler:2017knj}. 
Our calculation shows that a correction of the running weak 
mixing angle at zero momentum is not needed if its definition 
is based on the well-defined framework of the LEFT where 
all couplings are determined in such a way that predictions 
for matrix elements at low-energy agree with the SM calculation.

\section{Conclusion}
\label{sec:Discussion}

We have presented an explicit calculation of the matching 
condition that relates the weak mixing angle in the 
Standard Model and the Low Energy Effective Field Theory. 
This relation should be used to match the solutions of 
the renormalization group equation of the weak mixing 
angle below and above the $W$-boson threshold. In this 
way, the running weak mixing angle can be used in well-defined 
theories for all energy scales. 

The running weak mixing angle by itself is, of course, not 
sufficient to obtain predictions for observables with high 
precision. Additional higher-order corrections not absorbed 
in the running couplings have to be calculated. In principle, 
any prescription could be used to define the running weak 
mixing angle. However, a well-defined theoretical framework 
should be preferred. We therefore hope that our result 
provides the prerequisite for consistent and possible 
simplified calculations of higher-order corrections 
which will be relevant in the future when experimental 
results with high accuracy become available.

\section*{Acknowledgment}

We thank Jens Erler, Rodolfo Ferro-Hernandez and Mikhail 
Gorchtein for a careful reading of the manuscript and for 
comments which helped us to improve the presentation of 
our calculation. This work was supported by the Deutsche 
Forschungsgemeinschaft (DFG) in the framework of the 
collaborative research center SFB1044 ``The Low-Energy 
Frontier of the Standard Model: From Quarks and Gluons to 
Hadrons and Nuclei''.



\end{document}